\documentclass[12pt,preprint]{aastex}

\shorttitle{Importance of binaries in spectral studies}

\shortauthors{Zhongmu Li}

\begin{document}
\title{Potential Importance of Binary Evolution in UV-Optical Spectral Fitting of Early-Type Galaxies}
\author{Zhongmu Li$^{1,2,3}$, Caiyan Mao$^{1}$, Li Chen$^{1}$, Qian Zhang$^{1}$, Maocai Li$^{1}$}

\affil{$^{1}$Institute for Astronomy and History of Science and
Technology, Dali University, Dali 671003,
   China; \email{zhongmu.li@gmail.com}}
\affil{$^{2}$Astonomisches Rechen-Institut, Zentrum f\"{u}r
Astronomie der Universit\"{a}t Heidelberg, 69120 Heidelberg,
  Germany}
\affil{$^{3}$National Astronomical Observatories, Chinese Academy of
Sciences, Beijing 100012, China}

\begin{abstract}
Binaries are very common in galaxies, and more than half of Galactic
hot subdwarf stars, which are thought as a possible origin of
UV-upturn of old stellar populations, are found in binaries.
Previous works showed that binary evolution can make the spectra of
binary star populations significantly different from those of single
star populations. However, the effect of binary evolution has not
been taken into account in most works of spectral fitting of
galaxies. This paper studies the role of binary evolution in
spectral fitting of early-type galaxies, via a stellar population
synthesis model including both single and binary star populations.
Spectra from ultraviolet to optical band are fitted to determine a
few parameters of galaxies. The results show that the inclusion of
binaries in stellar population models may lead to obvious change in
the determination of some galaxy parameters and therefore it is
potentially important for spectral studies. In particular, the ages
of young components of composite stellar populations become much
older when using binary star population models instead of single
star population models. This implies that binary star population
models will measure significantly different star formation histories
(SFHs) for galaxies compared to single star population models. In
addition, stellar population models with binary interactions measure
larger dust extinctions than single star population models on
average. It suggests that when using binary star population models
instead of single star population models, negative extinctions are
possibly unnecessary in spectral fitting of early-type galaxies.

\end{abstract}

\keywords{Galaxies: elliptical and lenticular, cD--- Galaxies:
stellar content--- Galaxies: star formation.}

\section{Introduction}
Binaries are common in galaxies, and their evolution is often
different from single stars if binary components are not too far
\citep{Eggleton:2006}. In our Galaxy, the fraction of binaries is
shown as large as about 50\%. Binaries are found in different parts
of the Galaxy, e.g., Galactic field, halo, and thick and thin discs
\citep{1991AA...248..485D,1992ApJ...396..178F,1992ASPC...32...73M,
2010ApJS..190....1R,2005AJ....129.1886C}. In addition, most stars
massive than the sun are born in binaries
\citep{2010IAUS..262....3B}, and the massive the stars, the larger
the fraction of binaries \citep{2003ARAA..41...57L}. Therefore,
binaries are absolutely important for understanding the stellar
populations of galaxies, and it is necessary to study the role of
binary evolution in stellar population studies. The importance of
binaries has been recently listed as one of six important challenges
in stellar population studies in the next decades by
\cite{2010IAUS..262....3B}. However, most previous works used single
stars for modeling stellar populations because of the complexity of
binary evolution and more input parameters in binary star stellar
population synthesis (bsSPS). Besides, astronomers hope that binary
evolution does not change the results of single star stellar
population (ssSP) studies too much. If this is true, to take single
star stellar population synthesis (ssSPS) will be a good choice for
stellar population studies.

Meanwhile, a few works (e.g.,
\citealt{2006ChJAA...6..669L,2007MNRAS.380.1098H,2008MNRAS.387..105L,
2008IAUS..252..359L,2008ApJ...685..225L,2012MNRAS.424..874L,
2011RMxAC..40..277H}) have tried to investigate the effects of
binary evolution on stellar population studies and have used binary
star stellar population (bsSP) model in a few works (e.g.,
\citealt{2006ChJAA...6..669L,2010RAA....10..135L,2012ApJL,2011MNRAS.417.1702M,2012AA...543A...8M}).
It shows that binary evolution is able to reproduce many interesting
results such as blue stragglers, red stragglers, extended main
sequence \citep{2008MNRAS.387..105L, 2012ApJL}, and UV flux excess
(hereafter UV-upturn) in elliptical galaxies (e.g.,
\citealt{2007MNRAS.380.1098H, 2008MNRAS.387..105L,
2012MNRAS.424..874L}), without any special assumptions like very old
age or high mass-loss (see
\citealt{1994ApJ...430L.113L,1997ApJ...476...28P,1994ApJS...94...63B,1997ApJ...482..677Y}
for comparison). Binary evolution is shown as a natural explanation
for UV-upturn of elliptical galaxies, because more than half of
Galactic hot subdwarf stars are found in binaries
\citep{1984ApJ...287..320F,1994AJ....107.1565A,1995A&A...303..773T,
1998A&AS..132....1U,2001A&A...368..994A,2001MNRAS.326.1391M,2001PASP..113..944W,2004PASP..116..506R},
and it is well established that the vast majority (and quite
possibly all) of hot subdwarfs are the results of binary
interactions \citep{2007MNRAS.380.1098H}. On the other side, the
widely used ssSP models fail to reproduce UV-upturn for old stellar
populations under normal assumptions. Thus it suggests that binaries
are important and possibly the main contributors to UV-upturn
spectra of old populations. In addition, it shows that binaries can
change the results of simple stellar population (SSP) studies
obviously \citep{2008ApJ...685..225L}, when spectral line indices or
colours are used to determine stellar population parameters (age and
metallicity) of galaxies. Although the importance of binaries has
been shown, it is far from well understanding the role of binaries.
Deeper and more detailed studies are therefore needed. Because it is
shown that binary evolution leads to significant change in the UV
and optical spectra of populations \citep{2012MNRAS.424..874L}, and
fitting such spectra are the basis of most stellar population
studies (see, the review of \citealt{2011Ap&SS.331....1W} or papers
such as \citealt{2004MNRAS.355..273C,
2005MNRAS.358..363C,2006AJ....131..790L,2007MNRAS.381.1252T,
2008MNRAS.385.1998K,2009ApJS..185....1T, 2009A&A...501.1269K}), it
is important to study the effect of binaries on full spectrum
fitting.

This work aims to investigate the potential importance of binary
evolution in spectral studies of early-type galaxies. The reason for
studying early-type galaxies is that their stellar populations are
relatively simple, which can possibly be modeled by two population
components \citep{2012MNRAS.424..874L}. Our main purpose is to study
how the results of spectral fitting can be changed by using bsSP
models instead of ssSP models. UV-optical spectra will be fitted by
homogeneous ssSP and bsSP models that vary from only including
binary interactions or not when evolving stars.

The structure of this paper is as follows. Sect. 2 introduces
stellar population models, and Sect. 3 describes spectral synthesis.
Sect. 4 then presents the spectral fitting code (BS2fit) used in
this work. In Sect. 5, for some mock galaxies, the galaxy parameters
determined by ssSP and bsSP models are compared. Then in Sect. 6, a
sample of 10 early-type galaxies are used to give a similar test.
Finally, Sect. 7 summarizes and discusses on the results.

\section{Stellar population model}

An advanced stellar population synthesis model including both ssSPs
and bsSPs \citep{2008MNRAS.387..105L,2012MNRAS.424..874L} is taken
for this work. This model supplies homogeneous spectra for both
simple and composite stellar populations (SSPs and CSPs) with or
without binary interactions. This makes it possible to study the
result changes caused only by binary evolution. In the following two
subsections, we introduce the stellar population synthesis model for
SSPs and CSPs, respectively.

\subsection{Simple stellar populations}
Simple stellar populations (SSPs) are built as follows. Firstly,
some stars (primary components of binaries) are generated following
an initial mass function (IMF) of \citet{Chabrier:2003} with lower
and upper mass limits of 0.1 and 100 M$_\odot$ respectively. Then
for each binary, the mass of its secondary component is generated by
taking an uniform distribution for the mass ratio ($q$, 0--1) of
secondary to primary component (\citealt{Mazeh:1992};
\citealt{Goldberg:1994}). The separation ($a$) between two binary
components is given following an assumption that the fraction of
binary in an interval of log($a$) is constant when $a$ is big
(10$R_\odot$ $< a <$ 5.75 $\times$ 10$^{\rm 6}$$R_\odot$) and it
falls off smoothly when $a$ is small ($\leq$ 10$R_\odot$)
\citep{Han:1995}, which can be written as
\begin{equation}
  a~.p(a) = \left\{
            \begin{array}{ll}
            a_{\rm sep}(a/a_{\rm 0})^{\psi}, &~a \leq a_{\rm 0}\\
            a_{\rm sep}, &~a_{\rm 0} < a < a_{\rm 1}\\
     \end{array}
    \right.
\end{equation}
where $a_{\rm sep} \approx 0.070, a_{\rm 0} = 10R_{\odot}, a_{\rm 1}
= 5.75 \times 10^{\rm 6}R_\odot$ and $\psi \approx 1.2$. The
eccentricity ($e$) of each binary system is generated according to
an uniform distribution. In this method, each population contains
about 50\% binaries with orbital period less than 100\,yr, which is
similar to the typical binary fraction of the Galaxy. When building
binary star simple stellar populations (bsSSPs), the interactions
between two binary components are taken into account when evolving
stars, but all stars are evolved separately when building single
star simple stellar populations (ssSSPs). All stars of an SSP are
assumed to form in a star burst and have the same metallicity.

After the generation of population stars, all of them (4\,000\,000)
are evolved using the rapid stellar evolution code of
\citet{2002MNRAS.329..897H} (Hurley code). Most binary evolution
processes such as mass transfer, mass accretion, common-envelope
evolution, collisions, supernova kicks and angular momentum loss are
included for bsSPs. Different mass transfer mechanisms, i.e.,
dynamical mass transfer, nuclear mass transfer and thermal mass
transfer are taken into account using the results of many works
(e.g., \citealt{1997MNRAS.291..732T,1987ApJ...318..794H}). One can
see \cite{2002MNRAS.329..897H} for more details. Some default values
for Hurley code, i.e., 0.5, 1.5, 1.0, 0.0, 0.001, 3.0, 190.0, 0.5,
and 0.5, are taken for wind velocity factor ($\beta_{\rm w}$),
Bondi-Hoyle wind accretion faction ($\alpha_{\rm w}$), wind
accretion efficiency factor ($\mu_{\rm w}$), binary enhanced mass
loss parameter ($B_{\rm w}$), fraction of accreted material retained
in supernova eruption ($\epsilon$), common-envelope efficiency
($\alpha_{\rm CE}$), dispersion in the Maxwellian distribution for
the supernovae kick speed ($\sigma_{\rm k}$), Reimers coefficient
for mass loss ($\eta$), and binding energy factor ($\lambda$),
respectively. We take these default values, because they have been
checked in the work of \cite{2002MNRAS.329..897H}. Although these
default values remain somewhat large uncertainties, the results for
spectral stellar population synthesis will be not affected too much
by the uncertainties in these parameters, according to a test in our
previous work \citep{2012MNRAS.424..874L}. In addition, because
Hurley code uses some fitting formulae to calculate the evolution of
stars, it causes about 5\% uncertainty in the evolutionary
parameters of stars.

The evolutionary parameters of stars are finally transformed to the
spectral energy distributions (SED) (or spectra) of stellar
populations by BaSeL 3.1 spectral library \citep{Lejeune:1997,
Lejeune:1998, Westera:2002}. The library is chosen here because of
its wide wavelength coverage and reliability. The uncertainties in
the spectra of populations caused by the spectral library is small
(about 3\% on average). In the same way, the homogenous spectra of
both ssSSPs and bsSSPs are computed. Note that the only difference
between two kinds of models is that bsSSPs take binary interactions
into account but ssSSPs do not.

When comparing the spectra of bsSSPs to those of ssSSPs, we find
obvious difference. The left panels of Fig. 1 show the comparison of
spectra of two kinds of populations. It is clear that the spectra of
bsSSPs and ssSSPs are different, especially in UV band. Many old
($>$ 3\,Gyr, bottom lines) bsSSPs show UV-upturn spectra, which is
mainly caused by hot subdwarf and blue straggler stars (e.g.,
\citealt{2007MNRAS.380.1098H}), but all old ssSSPs do not have
similar spectra. Because UV-upturn phenomenon has been observed in
many elliptical galaxies \citep{1999ARAA..37..603O}, which are
usually thought as SSPs, our result suggests that bsSSPs can better
fit to early-type galaxies than ssSSPs. In binary evolution, the
formation channels for hot subdwarfs are common-envelope ejection,
stable Roche lobe overflow, and merger of helium white dwarfs (WDs)
(see also \citealt{2007MNRAS.380.1098H}).

\subsection{Composite stellar populations}
The SEDs of CSPs are built on the basis of SEDs of SSPs. Because
there is no common result for the star formation histories (SFHs) of
early-type galaxies, a simple method is taken to model CSPs. In
detail, each CSP is assumed to contain a pair of old and young
components with the same metallicity. This assumption is in
agreement of previous studies that early-type galaxies are dominated
by old populations and there is only a little fraction of young
populations in such galaxies. The mass fraction of young component
is assumed to be dependent on the ages of two components of CSPs,
which is calculated by formula (2). It means that the mass fraction
of young component declines exponentially with increasing difference
between the ages of old and young components. This agrees with
previous studies on the SFHs of early-type galaxies, e.g.,
\citet{Thomas:2005}.
\begin{equation}
  F_{\rm 2} = 0.5 ~{\rm exp}[\frac{t_{\rm2}-t_{\rm1}}{\rm
  \tau}]
\end{equation}
where $F_{\rm 2}$ is the mass fraction of young component;
$t_{\rm1}$ and $t_{\rm2}$ are the ages of old and young components
of a CSP, respectively. As a standard model, $\tau$ is taken as
3.02, according to the observational fraction of bright early-type
galaxies with recent ($\leq$ 1\,Gyr) star formation at a level more
than 1--2\% \citep{Yi:2005,2007A&A...471..795L}.

By the above method, the SEDs of single and binary star composite
stellar populations (ssCSPs and bsCSPs) are calculated. In the right
panels of Fig. 1, the SEDs of two kinds of CSPs are compared. For
some pairs of ssCSP and bsCSP with the same parameters, we see clear
difference between the SEDs of two kinds of populations. Similar to
the case of SSPs, the UV spectra of two kinds of populations are
obviously different. In addition, we find that both ssCSPs and
bsCSPs can show UV-upturn spectra, but UV-upturn comes from
different reasons. The UV-upturn of bsCSPs are mainly caused by
binary evolution, but that of ssCSPs by young stars. This suggests
that bsCSP and ssCSP models possibly give different estimates for
the SFHs of early-type galaxies.

\begin{figure} 
\includegraphics[angle=-90,width=160mm]{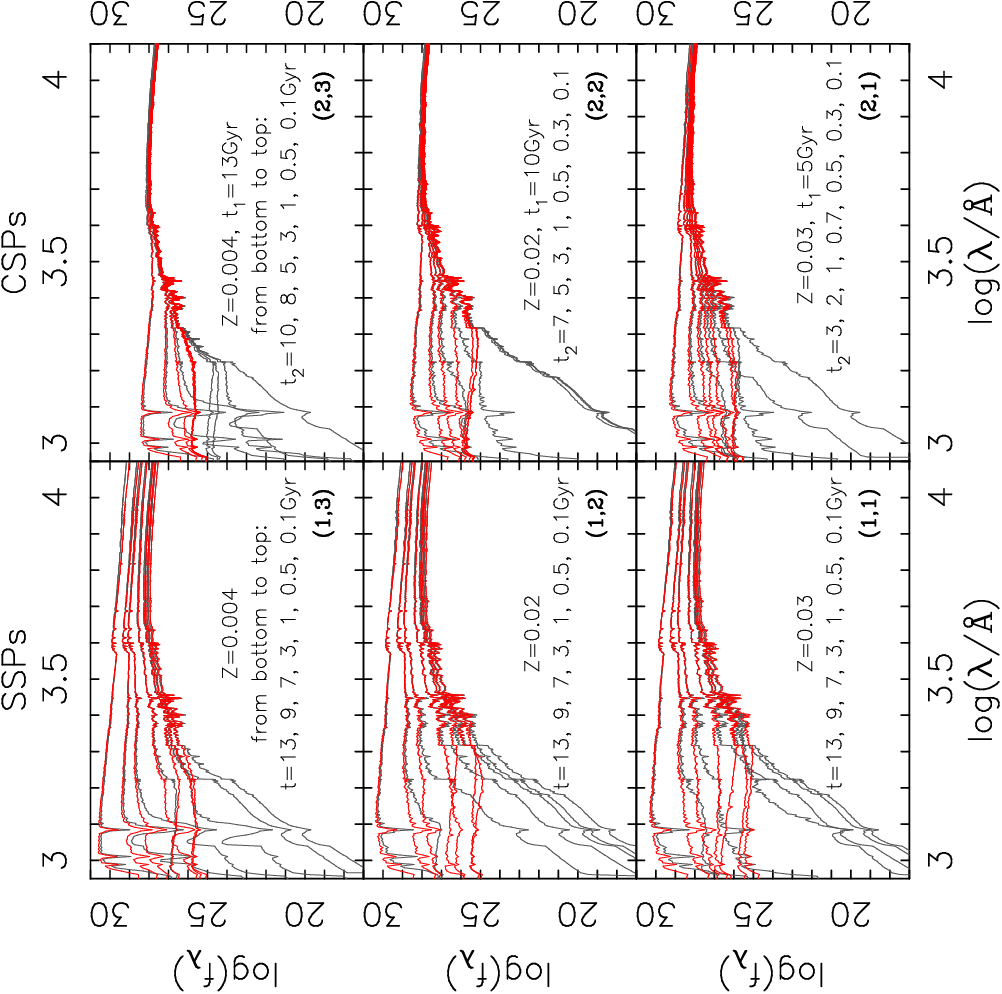}
\caption{Comparison of SEDs of single and binary star
populations. Grey lines are for single star populations, while red
lines for binary star populations. Left and right columns show the
results of simple and composite stellar populations (SSPs and CSPs),
respectively. Note that stellar age increases from top to bottom in
each panel.}
\end{figure}

\section{Spectral synthesis}
The observed spectra of stellar populations are built based on the
SEDs of CSPs. According to our model, each CSP is described by three
parameters: metallicity $Z$, old-component age $t_{\rm 1}$, and
young-component age $t_{\rm 2}$. The effects of stellar velocity
dispersion $\sigma_*$, extinction from dust around stellar
population, redshift, and the extinction caused by the Milky Way are
added to the SEDs of CSPs to form the observed spectra. Similar to
the works of \cite{2005MNRAS.358..363C}, the Line-of-sight stellar
motions are modeled by a Gaussian distribution $G$ centered at
velocity $v_*$ and with dispersion $\sigma_*$. The effect of dust
around population, which is parametrized by $V$-band optical depth
$\tau_{\rm{V}}$, is modeled following the work of
\cite{2000ApJ...539..718C}. When modeling the Galactic extinction,
which can be parametrized by $V$-band extinction $A_{\rm{V}}$, the
extinction law of \cite{1989ApJ...345..245C} with $R_{\rm{V}}$ = 3.1
is adopted. The observed flux ($f_\lambda$) at wavelength $\lambda$
can be expressed by
\begin{equation}\label{eq1}
f_{\lambda}=[f_{\lambda0}(z, t_{\rm{1}}, t_{\rm{2}}) \otimes G(v_*,
\sigma_*)] ~p(\tau_{\lambda}) ~r(A_{\lambda}),
\end{equation}
where $f_{\lambda0}$ is the flux of a CSP with three parameters ($Z,
t_{\rm{1}}, t_{\rm{2}}$), $G(v_*$, $\sigma_*$) is a Gaussian
distribution with mean and standard deviation of $v_*$ and
$\sigma_*$ respectively, $p(\tau_{\lambda})$ is the percentage of
energy that pass through the dust with optical depth of
$\tau_{\lambda}$, $r(A_{\lambda}$) is the fraction of energy after
the extinction ($A_{\lambda}$) of Milky Way galaxy. $\otimes$
denotes the convolution operator.

Fig. 2 gives two examples for the synthesis of the spectra of a pair
of ssCSP and bsCSP. We see that Line-of-sight stellar motions
(redshift and stellar velocity dispersion) affect the wavelength
range and strength of absorption lines obviously, but stellar
velocity dispersion does not affect continuum spectra too much.
Meanwhile, the dust around stellar population and the Milky Way
affects the whole spectra. We can also see significant difference
between the spectra of ssCSPs and bsCSPs from this figure.

\begin{figure} 
\includegraphics[angle=-90,width=\textwidth]{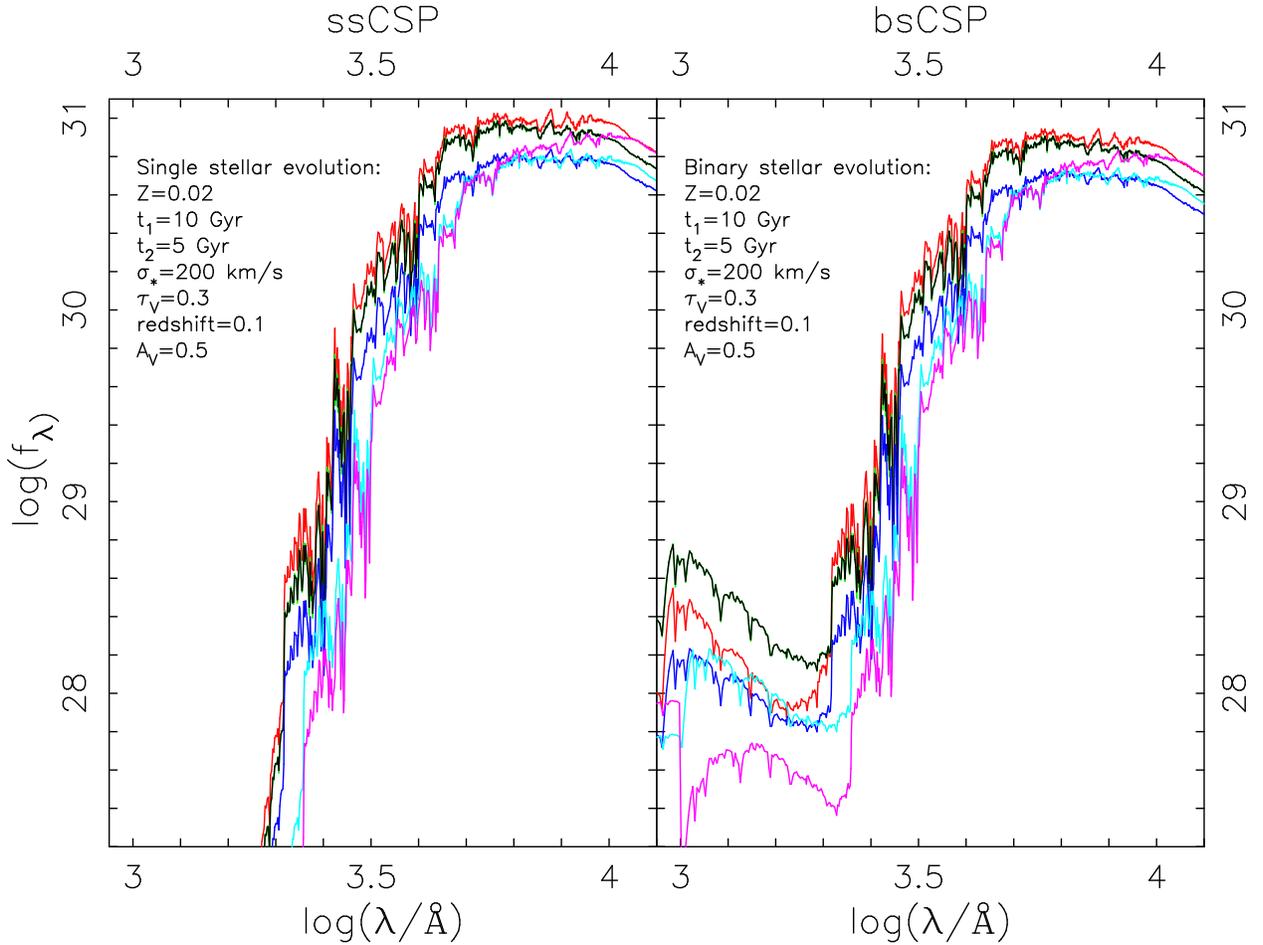}
\caption{Formation of synthetic SEDs of single (left) and binary
(right) star composite stellar populations (ssCSP and bsCSP). Green
lines show SEDs of two simple stellar populations (SSPs) with
age = 10\,Gyr and solar metallicity ($Z$ = 0.02), while black, red,
blue, cyan, and purple lines are SEDs after adding the effects of
second star formation, stellar velocity dispersion, dust in galaxy,
redshift, and Galactic extinction into the original SEDs of two
SSPs, gradually.}
\end{figure}

\section{Spectral fitting: BS2fit code}
Spectral fitting is an important technique to determine many
properties of galaxies, e.g., redshift, stellar metallicity, age,
velocity dispersion, mass, and dust extinction, because different
wavelengths are usually dominated by various physical processes.
Thus it is a good choice to study the role of binary evolution in
stellar population studies via spectral fitting. Although there are
some available spectral fitting codes, e.g., MOPED
\citep{2000MNRAS.317..965H}, PLATEFIT \citep{2004ApJ...613..898T},
STARLIGHT (\citealt{2004MNRAS.355..273C, 2005MNRAS.358..363C}),
VESPA (\citealt{2007MNRAS.381.1252T, 2009ApJS..185....1T}), STECKMAP
\citep{2006MNRAS.365...74O}, sedfit \citep{2006ApJ...649..692W},
NBURSTS \citep{2007IAUS..241..175C}, and ULYSS
(\citealt{2009A&A...501.1269K}), they are not very suitable for this
work. Firstly, all of them are developed for some special works so
that their fitting methods and wavelength coverages are not suitable
for this work. For example, STARLIGHT aims to fit the optical
spectra of SDSS galaxies but our work needs a wider wavelength
coverage. VESPA studies the SFHs of galaxies via some binned
parameters, but we want to do more accurate studies. Furthermore,
there are some other limitations to use these codes, e.g., the
dependence of internet connection or unaltered procedures.
Therefore, we decide to create a new code. We call our new code
binary star to fit (hereafter BS2fit), which can be used for
spectral fitting via both bsSPs and ssSPs, within a wide wavelength
range from UV to optical band. The new code can be briefly
introduced as follows.

BS2fit code will be revised and updated\footnote{The code and data
can be obtained on request to Zhongmu Li, and the authors are trying
to make it available on the internet.}. At this stage, BS2fit aims
to determine a few parameters of early-type galaxies from observed
spectra using bsSP or ssSP models. An observed spectrum is assumed
to be determined by eight input parameters, i.e., $Z$, $t_{\rm{1}}$,
$t_{\rm{2}}$, $v_*$, $\sigma_*$, $\tau_{\rm{V}}$, $A_{\rm{V}}$, and
total stellar mass $M_*$. Because there are so many input
parameters, it is not easy to get a best-fit result by directly
searching for the combination of parameters with minimum $\chi^2$ in
a high-resolution parameter grid. Therefore, we divide the fitting
task into three parts. First, Line-of-sight velocity (or redshift)
is determined by a few methods that are widely used (e.g.,
absorption or emission line fitting and cross-correlation analysis)
(see e.g., \citealt{2001MNRAS.328.1039C}). Then the Galactic
extinction parameter $A_{\rm{V}}$ is determined from the direction
of galaxy and dust map of the Galaxy (e.g.,
\citealt{1982AJ.....87.1165B, 1998ApJ...500..525S}). Finally, the
other parameters are determined based on fixed redshift and Galactic
extinction. Although this has made fitting process much faster, it
is still difficult to get the result by a direct search on a
personal computer. Two techniques are therefore used to make the
fitting significantly quick. One is to estimate the possible ranges
of parameters using SSP models before comparing the observed
spectrum to those of CSPs. We call this procedure SSP-fitting. This
is very effective, because it guides the code to search in only
limited ranges of parameters. This method is also reliable for most
populations, because SSPs usually estimate lower metallicities,
smaller stellar velocity dispersions, and younger ages for
populations, compared to CSPs. In detail, Fig. 3 shows how the
ranges of parameters of 74 test stellar populations are related to
SSP-fitted parameters. On average, the parameter ranges of about 95
percent of CSPs can be well estimated from SSP-fitting.

The other technique for speeding up the fitting is to compare
continuum spectra before comparing whole spectra. This skips many
populations according to their continuum difference from the
observed spectra. This can be done because there is a maximum change
in continuum spectra corresponding to the effects of dust extinction
and stellar velocity dispersion when their ranges are given. The
goodness of a fit is judged by $\chi^2 = \Sigma [(f_{\rm ob\lambda}
- f_{\rm th\lambda})^2~\omega_{\lambda}]$, where $f_{\rm ob\lambda}$
and $f_{\rm th\lambda}$ are observed and theoretical spectra, while
$\omega_{\lambda}$ is the weight for wavelength $\lambda$. The
best-fit result corresponds to minimum $\chi^2$.

We mask out some emission line regions as they can affect the final
results obviously. According to some previous works
(\citealt{1995ApJS...97..331M, 2007MNRAS.378.1550P,
2007MNRAS.381.1252T, 2009ApJS..185....1T}), the following
emission-line regions in every spectrum's rest frame wavelength
range are masked out: 1213--1219, 1548--1550, 1637--1643,
1905--1911, 2323--2329, 2421--2427, 2796--2803, 3423--3429,
3711--3741, 4087--4117, 4325--4355, 4846--4876, 4944--4974,
4992--5022, 5885--5900, 6535--6565, 6548--6578, 6569--6599,
6702--6732, 6716--6746, 6728--6734, 7132--7138, 7319--7330 ${\rm
\AA}$.

Because we want to make the results accurate enough, a
high-resolution grid of parameters is taken. In detail, stellar
metallicity ($Z$) is from 0.0003 to 0.03, with an interval of 0.0001
when $Z$ is less than 0.001 and an interval of 0.001 for higher
metallicity. The old-component age ($t_{\rm 1}$) of populations is
in the range of zero to 15\,Gyr, with an interval of 0.1\,Gyr, and
the young-component age ($t_{\rm 2}$) decreases from $t_{\rm 1}$ to
zero. Stellar velocity dispersion, $\sigma_*$, changes from 0 to
350\,km s$^{-1}$ by a step of 10\,km s$^{-1}$, and the optical depth
($\tau_{\rm v}$) is between 0 and 1.5, with an interval of 0.01.
Note that the ranges of $\sigma_*$ and $\tau_{\rm v}$ are chosen
according to some previous works, e.g., \cite{2005MNRAS.358..363C}
and \cite{2007MNRAS.381.1252T}.

\begin{figure} 
\includegraphics[angle=-90,width=\textwidth]{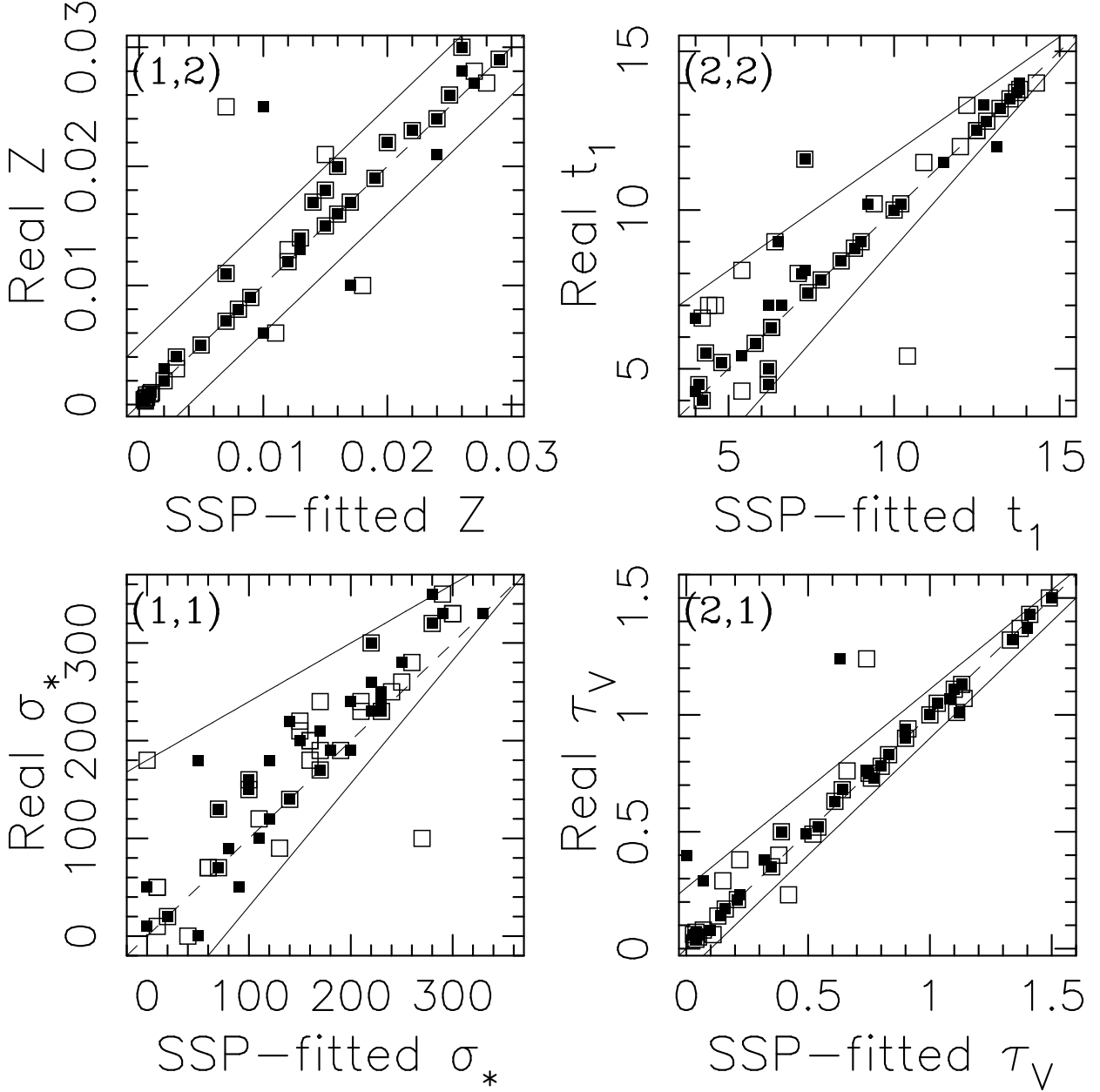}
\caption{Comparison of SSP-fitted and real parameters of 37 ssCSPs
and 37 bsCSPs. Open and filled squares are for ssCSPs and bsCSPs,
respectively. Solid lines show the ranges for searching the final
parameters via CSPs, while dashed lines are unity lines. $t_{\rm 1}$
is in Gyr and $\sigma_*$ is in km s$^{-1}$.}
\end{figure}

In order to check the accuracy of BS2fit code, Figs. 4 and 5 show
the input versus recovered values of mass-weighted age $<t_*>_{\rm
M}$, luminosity-weighted age $<t_*>_{\rm L}$, $Z$, $t_{\rm{1}}$,
$t_{\rm{2}}$, $v_*$, $\sigma_*$, $\tau_{\rm{V}}$, and $M_*$ of 37
ssCSPs (open squares) and 37 bsCSPs (filled squares). In the test
relating to Fig. 4 all wavelengths are assumed the same weight
because there is no error in spectra, while in the study
corresponding to Fig. 5 the weight of each wavelength is set to be
1/$E_{o\lambda}^2$, where $E_{o\lambda}$ is observational error. The
stellar mass of each population is calculated from the input flux at
5500\,${\rm \AA}$ and mass-to-light ratio of best-fit population. We
see that when there is no error in input spectra, BS2fit code can
well recover most parameters (Fig. 4), although the $t_{\rm{2}}$
values of some populations with large ($>$ 3.5\,Gyr) input
$t_{\rm{2}}$ are not recovered accurately because of the degeneracy
effects of various parameters on the same spectra. Note that the
input spectra of a pair of ssCSP and bsCSP with the same set of
input parameters are usually different.

If there are uncertainties (signal-to-noise ratio S/N = 10) in input
spectra (Fig. 5), BS2fit code reports obviously larger ages
(especially $t_{\rm 1}$) for most populations. This because the
change of spectra is less sensitive to the age change of old
populations. Moreover, the uncertainties in input spectra result in
lower estimates for stellar metallicities of some populations,
because the change of spectra has different sensitivities to the
metallicity changes of metal-poor and rich populations. The
deviations in stellar age and metallicity are also related to the
well-known age--metallicity degeneracy. Therefore, the S/N of
observed spectra is crucial for constraining the stellar
metallicities and ages of galaxies. In addition, we see that usually
bsCSP and ssCSP models lead to similar difference between the input
and recovered parameters. This means that BS2fit code can be used to
study the difference between galaxy parameters, which are caused by
taking various (bsCSP and ssCSP) models, although the values of
parameters may be not so accurate. Note that the fitted parameters
and uncertainties are given statistically by taking the average and
1 $\sigma$ deviation of the results of fitting some perturbed
spectra that are generated randomly according to the observational
uncertainty (see also \citealt{2005MNRAS.358..363C,
2007MNRAS.381.1252T}).

\begin{figure} 
\includegraphics[angle=-90,width=0.65\textwidth]{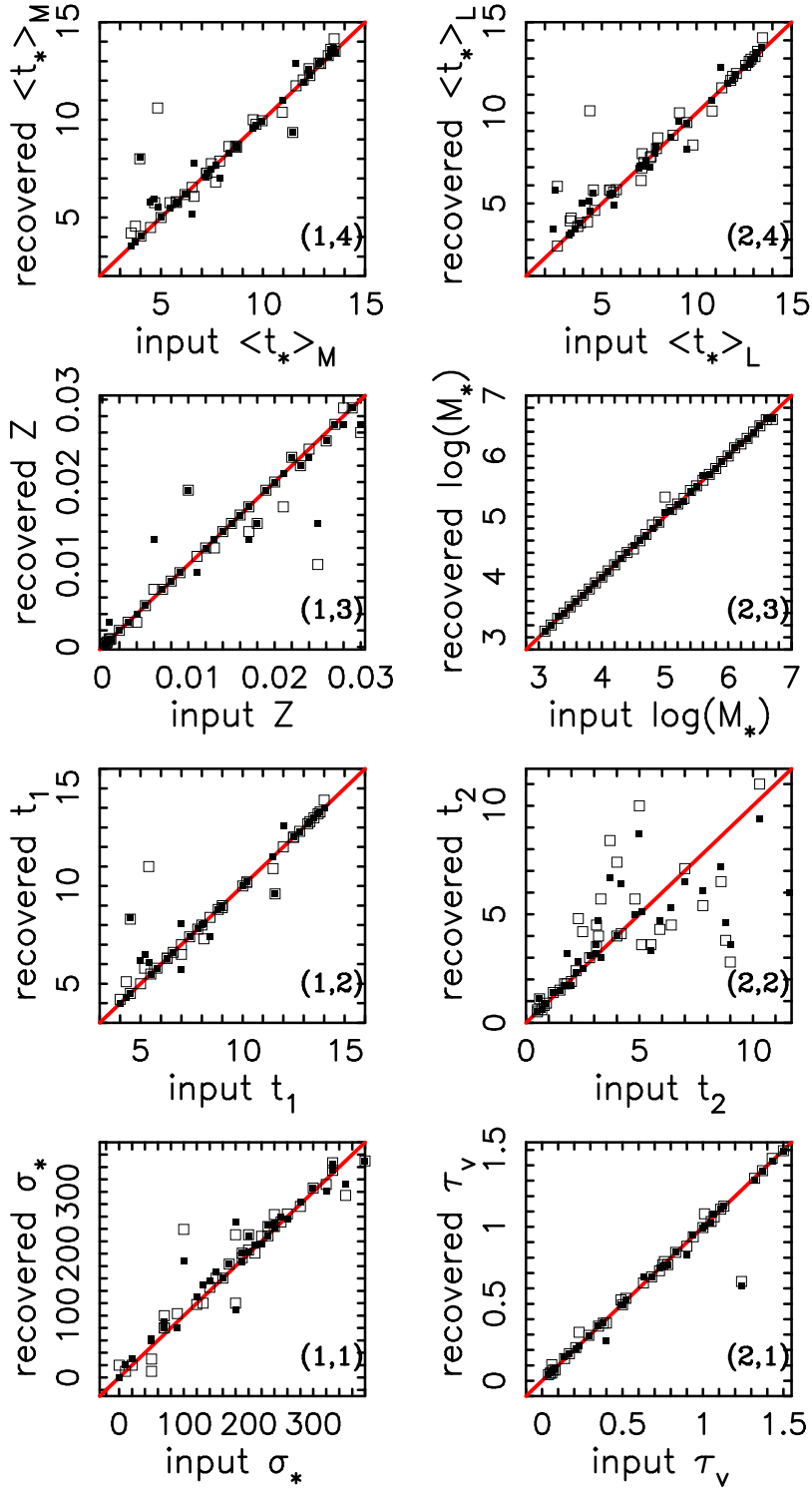}
\caption{Comparison of input and recovered parameters of ssCSPs and
bsCSPs. The input spectra have no error. Squares have the same
meanings as in Fig. 3. Solid line is a unity line. $<t_*>_{\rm M}$
and $<t_*>_{\rm L}$ denote mass and luminosity-weighted age,
respectively. $Z$, log($M_*$), $t_{\rm 1}$, $t_{\rm 2}$, $\sigma_*$
and $\tau_v$ represent metallicity, logarithmic stellar mass,
old-component age, young-component age, stellar velocity dispersion,
and optical depth. $<t_*>_{\rm M}$ and $<t_*>_{\rm L}$ are in Gyr,
and $M_*$ is in solar mass.}
\end{figure}

\begin{figure} 
\includegraphics[angle=-90,width=0.7\textwidth]{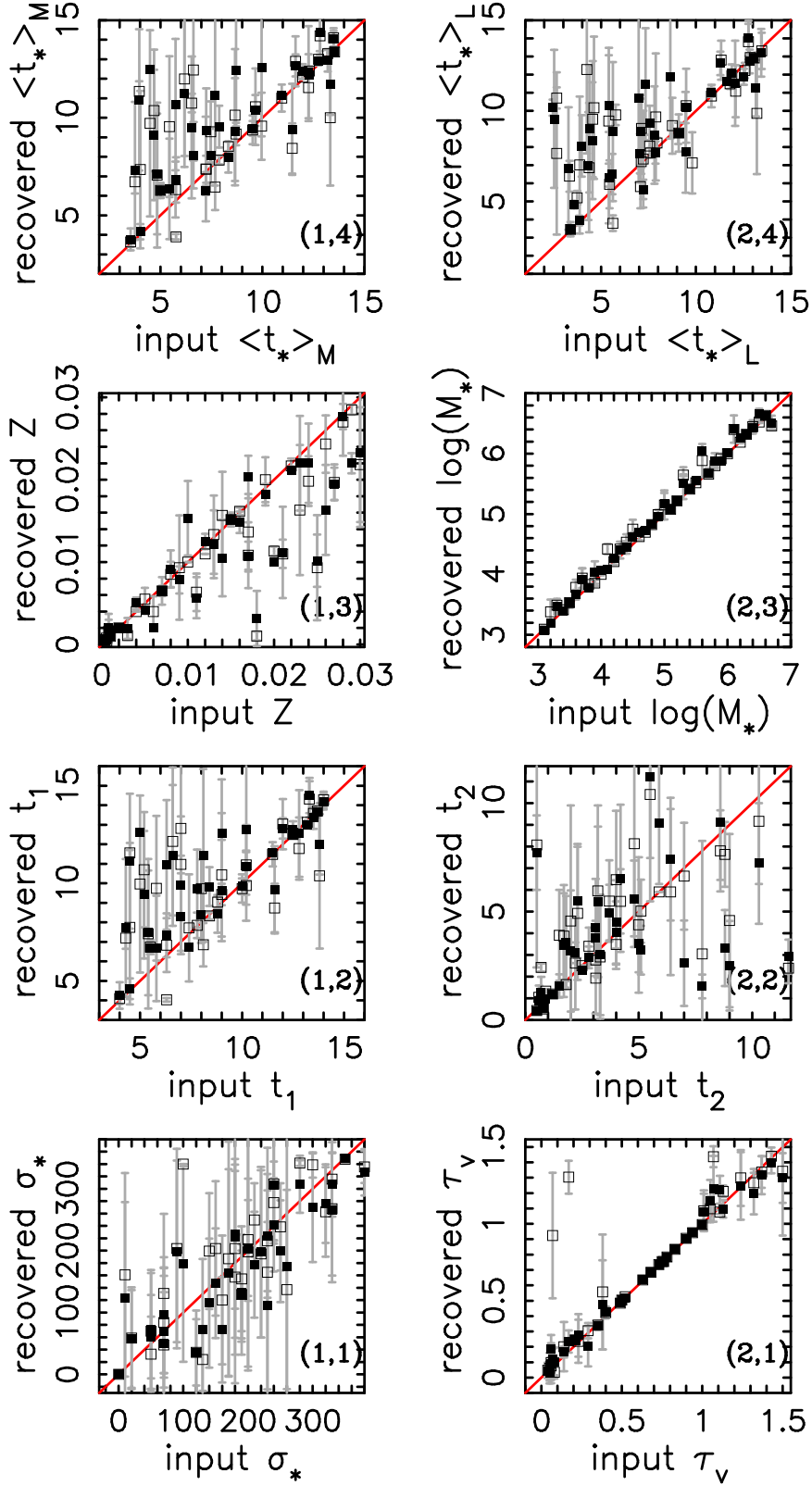}
\caption{Similar to Fig. 4, but for some input spectra with S/N =
10. Error bars show $\sigma$ uncertainties of galaxy parameters.}
\end{figure}

\section{Results from mock galaxies}
This section studies the difference between bsSP and ssSP-fitted
results, via a sample of 37 mock galaxies, which covers wide
parameter ranges. We assume that all mock galaxies are bsCSPs, as
almost all galaxies may contain some binaries. In addition, because
any observed spectrum possibly contains some uncertainties, the
effect of observational uncertainties is taken into account. Similar
to the previous section, the S/N of spectra is set to 10, and
$\omega_{\lambda}$ is given as 1/$E_{o\lambda}^2$.

Fig. 6 shows the comparison of bsSP and ssSP-fitted parameters of 37
bsCSPs (mock galaxies). We see that ssSP and bsSP-fitted results for
mass and luminosity-weighted ages, metallicity, stellar mass,
old-component age, stellar velocity dispersion, and galactic dust
extinction are roughly consistent when uncertainties are taken into
account. In particular, ssSP and bsSP-fitted stellar masses are
almost the same. However, bsSP-fitted young-component ages ($t_{\rm
2}$) are different from ssSP-fitted results, significantly. For mock
galaxies with ssSP-fitted $t_{\rm 2}$ younger than 3.5\,Gyr, bsSP
models measure much older young population components for most mock
galaxies. This suggests that ssSP models usually measure more recent
star formations for early-type galaxies, compared to bsSP models. In
this case, maybe many early-type galaxies do not contain so much
recent star formations as the results of \cite{2012MNRAS.421..314C},
which was obtained using some ssSP models.

\begin{figure} 
\includegraphics[angle=-90,width=0.7\textwidth]{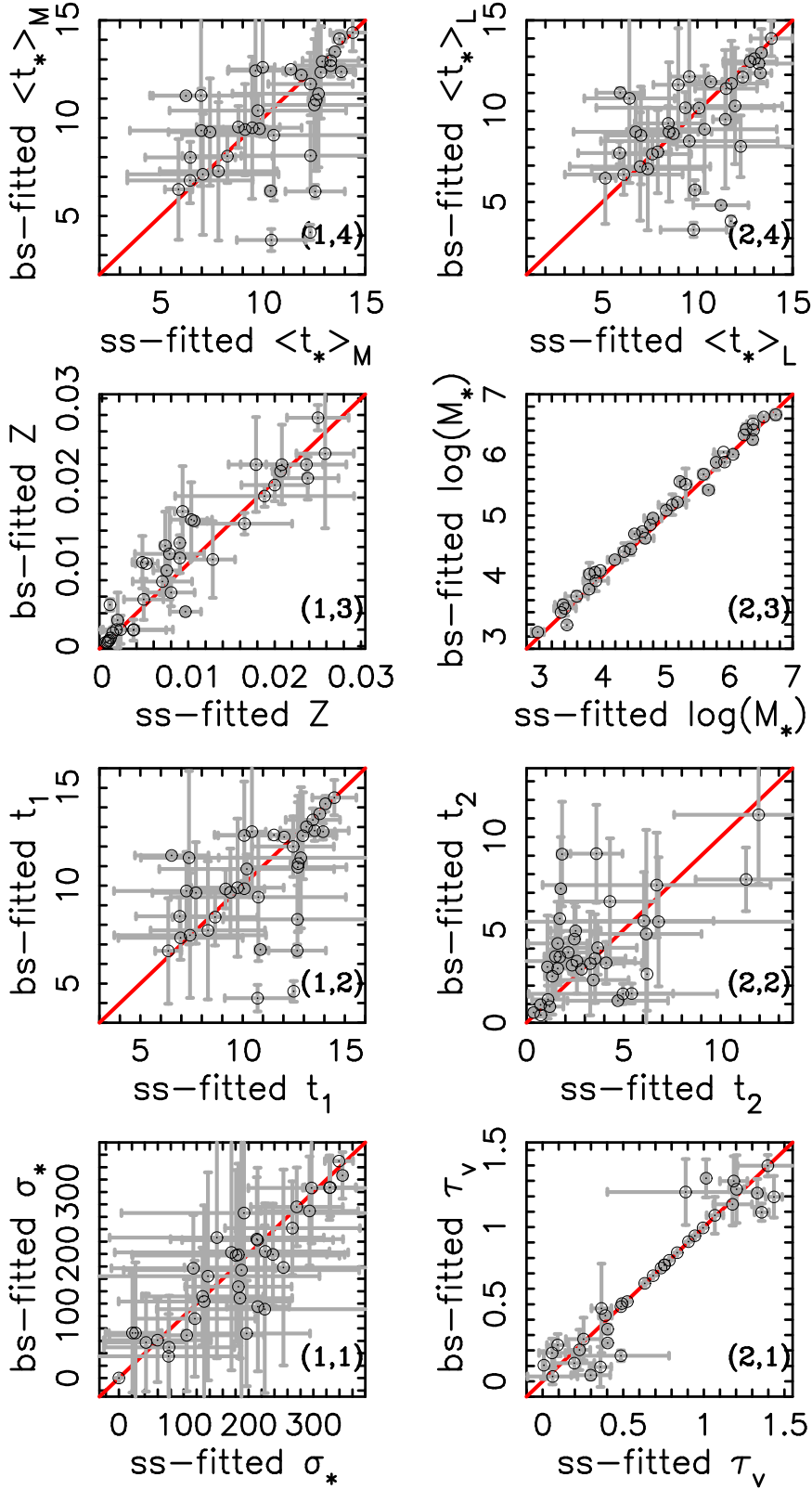}
\caption{Comparison of ssSP and bsSP-fitted parameters of 37 bsCSPs.
Points and error bars show mean parameters and $\sigma$
uncertainties, respectively. ``ss-fitted'' and ```bs-fitted'' denote
ssSP-fitted and bsSP-fitted results, respectively.}
\end{figure}

\section{Results from real galaxies}
In this section, we try to compare the galaxy parameters that are
determined via ssSP and bsSP models, for 10 early-type galaxies. The
data are selected from a catalogue of UV-optical spectra of 99 local
galaxies, which is available on the
internet\footnote{ftp://ftp.stsci.edu/pub/catalogs/nearby$_{-}$gal/sed.html}.
The advantage of using these data is the wide wavelength coverage
from UV to optical band. Because binaries affect UV spectra
obviously \citep{2007MNRAS.380.1098H, 2008MNRAS.387..105L,
2012MNRAS.424..874L}, such data seems ideal for exploring the effect
of binary evolution on spectral fitting. Besides some emission
lines, we also mask out the spectra around two conjunction
wavelengths (2000 and 3000 ${\rm \AA}$) as the two parts contain
obviously larger uncertainties. In addition, because the quality of
optical spectra is better than UV spectra, optical spectra are given
higher weights by taking S/N = 8, while the S/N of UV spectra is
taken as 7. The S/N values are given according to a description of
the data \citep{1996ApJ...467...38K}, but the data quality seems not
as good as S/N = 8, because the minimum $\chi^2$ values of fittings
are obviously larger than 1.

Using BS2fit code, six parameters of 10 galaxies are determined, and
the ssSP and bsSP-fitted results are then compared. For convenience,
we take the redshifts of 10 galaxies from NASA/IPAC Extragalactic
Database (NED). The original results were obtained by
 \cite{2003AJ....126.2268W,1991ApJS...75..935D,2008ApJ...676..184T,2004AJ....128...16K,
 1999ApJS..121..287H,2005MNRAS.356.1440D}. The Galactic
extinctions are calculated using the extinction calculator of NED,
which uses the data and technique of \cite{1998ApJ...500..525S}.
Tables 1 and 2 list the ssSP and bsSP-fitted parameters of 10
galaxies. Because the quality of observed spectra is not very good,
the result for stellar velocity dispersion may be not accurate. Fig.
7 compares the best-fit spectra of ssSP and bsSP fittings with
observed ones, and Fig. 8 compares ssSP and bsSP-fitted parameters.
We find that both ssSP and bsSP models can fit to the spectra of
eight galaxies, i.e, NGC1399, NGC1553, NGC2865, NGC3031, NGC1404,
NGC221, NGC1052, and NGC205. At the same time, both bsSP and ssSP
models do not fit the spectra part with wavelength larger than
6000\,${\rm \AA}$ for NGC210. This may result from the low upper
metallicity (0.03) of our theoretical populations. In addition, we
find that ssSPs can fit the observed spectra of NGC1433 better than
bsSPs. This implies that most stars of this galaxy may be single
stars, rather than binary stars.

Fig. 8 shows the comparison of ssSP and bsSP-fitted parameters of 10
galaxies. We see that ssSP and bsSP models report obviously
different results for many parameters of galaxies. In particular,
the bsSP-fitted young-component ages ($t_{\rm 2}$) of six galaxies
are significantly older than ssSP-fitted results, and the
bsSP-fitted dust extinctions ($\tau_v$) of half galaxies are
obviously larger than ssSP-fitted results. In addition, two kinds of
stellar population models give different results for the
mass-weighted, luminosity-weighted, and old-component ages of some
galaxies. Furthermore, we see some difference between the bsSP and
ssSP-fitted stellar velocity dispersions ($\sigma_*$). However, bsSP
and ssSP models measure similar stellar metallicities and masses for
most galaxies. When comparing the results of 10 real galaxies and 37
mock galaxies, we find good consistency: bsSP models measure older
young-component ages and larger dust extinctions than ssSP models,
while two types of models measure similar stellar metallicities and
masses for galaxies. Because BS2fit can recover the ages of recent
($t_{\rm 2} \leq$ 3.5\,Gyr) star bursts well (see Fig. 4), both
tests of mock and real galaxies indicate that binary evolution
affects the measurement of SFHs of early-type galaxies,
significantly.

\begin{figure} 
\includegraphics[angle=-90,width=\textwidth]{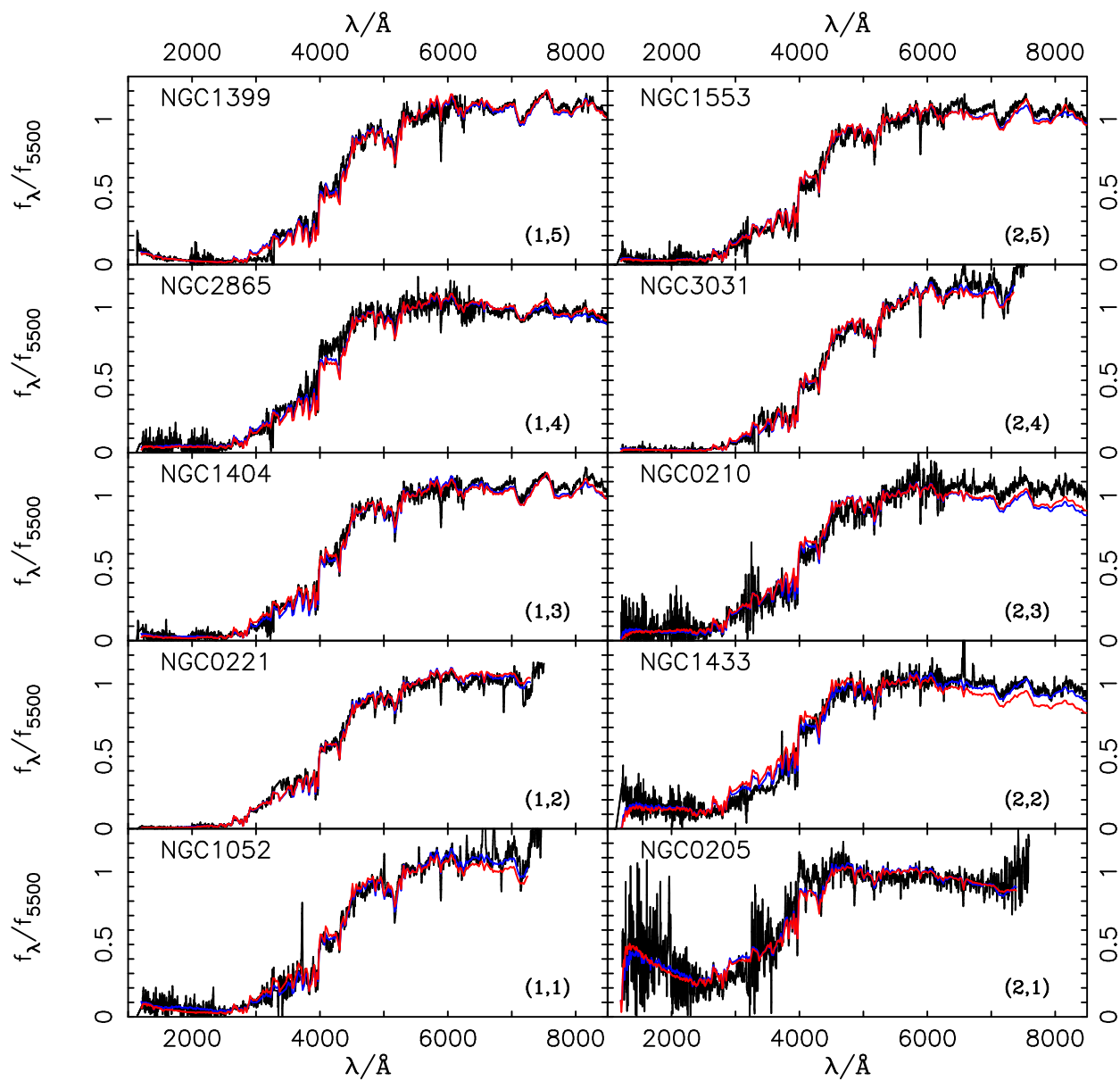}
\caption{Comparison of bsSP and ssSP-fitted SEDs with observational
spectra of 10 galaxies. Black, red and blue lines are for observed,
bsSP-fitted, and ssSP-fitted spectra, respectively. All spectra are
normalized at 5500\,${\rm \AA}$.}
\end{figure}

\begin{figure} 
\includegraphics[angle=-90,width=0.75\textwidth]{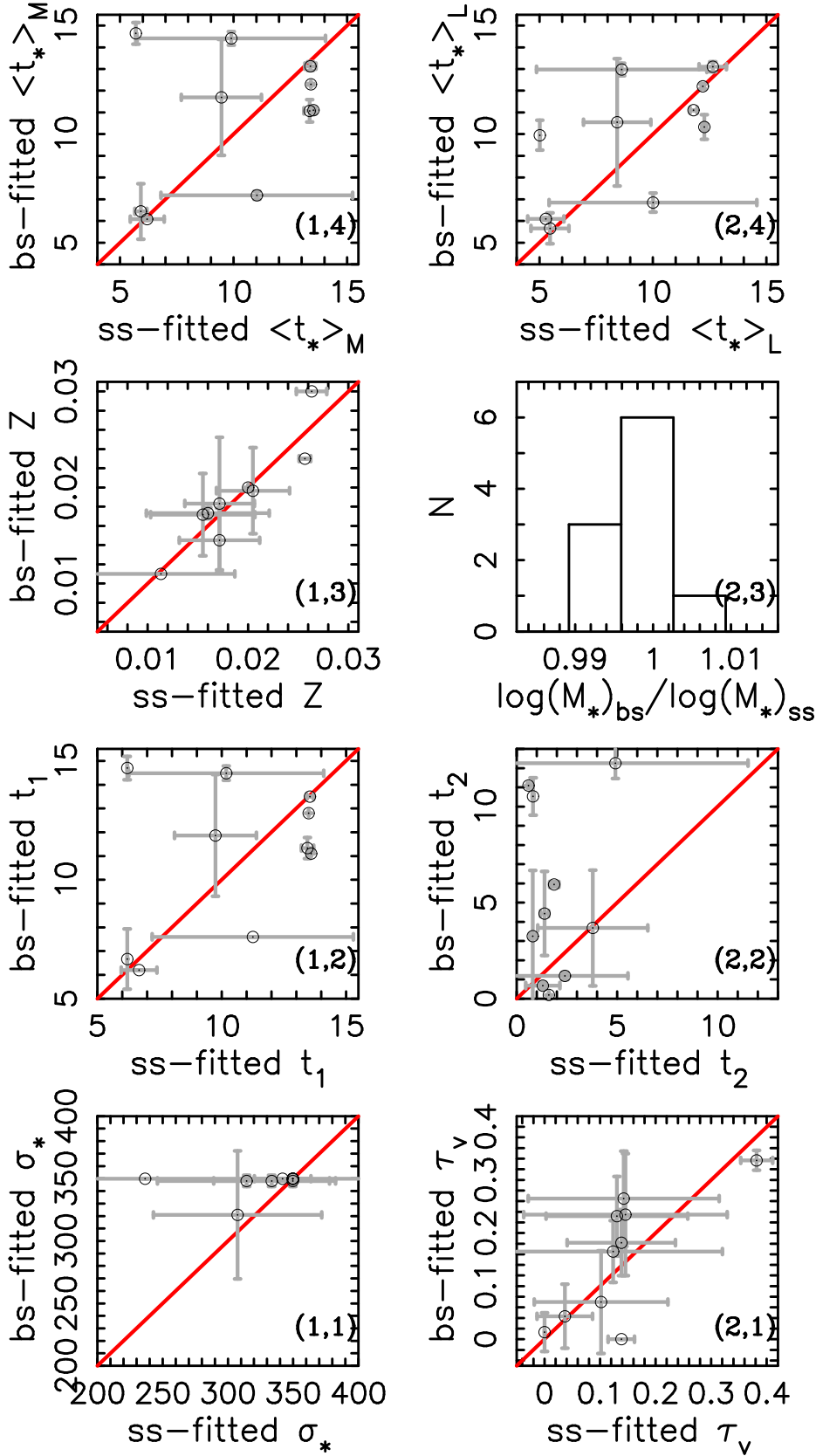}
\caption{Comparison of ssSP and bsSP-fitted parameters of 10 galaxies. Solid lines are unity lines.
log$(M_*)_{\rm bs}$ and log$(M_*)_{\rm ss}$
in panel (2,3) are bsSP and ssSP-fitted stellar mass, respectively.}
\end{figure}

\begin{deluxetable}{crrrrrrrrr}
\tabletypesize{\scriptsize} \tablecaption{ssSP-fitted parameters of
10 galaxies. The unit of $t_{\rm 1}$, $t_{\rm 2}$, $<t_{*}>_{\rm M}$
and $<t_{*}>_{\rm L}$ is Gyr, and that of $\sigma_{*}$ is km
s$^{-1}$. The values of redshift and $A_{\rm V}$ are taken from
NED.\label{tb1}} \tablewidth{0pt} \tablehead{ \colhead{Galaxy} &
\colhead{redshift} & \colhead{$A_{\rm V}$} & \colhead{$Z$} &
\colhead{$t_{\rm 1}$}  & \colhead{$t_{\rm 2}$} &
\colhead{$\sigma_{*}$} & \colhead{$\tau_{\rm V}$}  &
\colhead{$<t_{*}>_{\rm M}$}   & \colhead{$<t_{*}>_{\rm L}$} }
\startdata
NGC1399 &0.004753  &0.0393 &0.0263$\pm$0.0015 &13.55$\pm$0.14  &4.92$\pm$6.59 &237$\pm$195 &0.138$\pm$0.097 &13.38$\pm$0.24 &12.63$\pm$0.61\\
NGC1553 &0.004236  &0.0468 &0.0172$\pm$0.0035 &9.75$\pm$1.65   &2.40$\pm$3.13  &333$\pm$44  &0.143$\pm$0.172 &9.46$\pm$1.76 &8.42$\pm$1.49 \\
NGC2865 &0.008619  &0.1608 &0.0172$\pm$0.0040  &10.18$\pm$3.92 &1.30$\pm$0.85  &350$\pm$0  &0.037$\pm$0.050 &9.90$\pm$4.15 &8.62$\pm$3.75\\
NGC3031 &-0.000113 &0.1486 &0.0257$\pm$0.0006 &13.50$\pm$0.00  &0.82$\pm$0.05  &350$\pm$0  &0.382$\pm$0.029 &13.40$\pm$0.00 &12.19$\pm$0.06\\
NGC1404 &0.006391  &0.0353 &0.0205$\pm$0.0036 &13.45$\pm$0.25 &0.80$\pm$0.16  &314$\pm$68  &0.130$\pm$0.128 &13.35$\pm$0.26 &12.26$\pm$0.11\\
NGC 210 &0.005457  &0.0673 &0.0160$\pm$0.0061 &6.68$\pm$0.72 &1.87$\pm$0.23 &342$\pm$22 &0.146$\pm$0.183 &6.20$\pm$0.75 &5.28$\pm$0.80\\
NGC 221 &-0.000667 &0.4797  &0.0113$\pm$0.0074  &11.25$\pm$4.04 &1.38$\pm$0.20  &350$\pm$0  &0.124$\pm$0.197 &11.02$\pm$4.22 &10.00$\pm$4.57\\
NGC1433 &0.003586  &0.0282 &0.0155$\pm$0.0052 &6.20$\pm$0.00 &3.78$\pm$2.74 &350$\pm$0  &0.102$\pm$0.120 &5.91$\pm$0.26 &5.46$\pm$0.83\\
NGC1052 &0.001004 &0.0824 &0.0200$\pm$0.0000 &13.60$\pm$0.00 &0.58$\pm$0.05  &350$\pm$0  &0.138$\pm$0.024 &13.51$\pm$0.00 &11.79$\pm$0.15\\
NGC 205 &-0.000777 &0.1685  &0.0122$\pm$0.0005  &6.20$\pm$0.00 &1.60$\pm$0.00 &307$\pm$64 &0.000$\pm$0.000 &5.70$\pm$0.00 &5.02$\pm$0.01\\
 \enddata
\end{deluxetable}

\begin{deluxetable}{crrrrrrrrr}
\tabletypesize{\scriptsize} \tablecaption{Similar to Table 1, but
for bsSP-fitted parameters. \label{tb2}} \tablewidth{0pt}
\tablehead{ \colhead{Galaxy} & \colhead{redshift} & \colhead{$A_{\rm
V}$} & \colhead{$Z$} & \colhead{$t_{\rm 1}$}  & \colhead{$t_{\rm
2}$} & \colhead{$\sigma_{*}$} & \colhead{$\tau_{\rm V}$}  &
\colhead{$<t_{*}>_{\rm M}$}   & \colhead{$<t_{*}>_{\rm L}$} }
\startdata
NGC1399 &0.004753  &0.0393 &0.0300$\pm$0.0000 &13.50$\pm$0.00  &12.27$\pm$0.81 &350$\pm$0 &0.182$\pm$0.062 &13.13$\pm$0.20 &13.10$\pm$0.23\\
NGC1553 &0.004236  &0.0468 &0.0183$\pm$0.0069 &11.87$\pm$2.56   &1.18$\pm$0.19  &348$\pm$5  &0.265$\pm$0.090 &11.69$\pm$2.67 &10.55$\pm$2.93 \\
NGC2865 &0.008619  &0.1608 &0.0145$\pm$0.0033  &14.48$\pm$0.30 &0.68$\pm$0.16  &348$\pm$5  &0.043$\pm$0.060 &14.41$\pm$0.31 &12.98$\pm$0.29\\
NGC3031 &-0.000113 &0.1486 &0.0230$\pm$0.0000 &12.80$\pm$0.00  &10.53$\pm$0.97  &350$\pm$0  &0.337$\pm$0.019 &12.29$\pm$0.04 &12.20$\pm$0.11\\
NGC1404 &0.006391  &0.0353 &0.0197$\pm$0.0045 &11.33$\pm$0.45 &3.25$\pm$3.43  &348$\pm$5  &0.232$\pm$0.075 &11.07$\pm$0.52 &10.33$\pm$0.56\\
NGC 210 &0.005457  &0.0673 &0.0173$\pm$0.0006 &6.20$\pm$0.00 &5.95$\pm$0.14 &350$\pm$0 &0.235$\pm$0.115 &6.09$\pm$0.06 &6.09$\pm$0.06\\
NGC 221 &-0.000667 &0.4797  &0.0110$\pm$0.0000  &7.60$\pm$0.00 &4.43$\pm$2.19  &350$\pm$0  &0.165$\pm$0.058 &7.18$\pm$0.19 &6.85$\pm$0.44\\
NGC1433 &0.003586  &0.0282 &0.0172$\pm$0.0043 &6.67$\pm$1.27 &3.68$\pm$3.01 &350$\pm$0  &0.070$\pm$0.097 &6.44$\pm$1.28 &5.66$\pm$0.71\\
NGC1052 &0.001004 &0.0824 &0.0200$\pm$0.0000 &11.10$\pm$0.00 &11.10$\pm$0.00  &350$\pm$0  &0.000$\pm$0.000 &11.10$\pm$0.00 &11.10$\pm$0.00\\
NGC 205 &-0.000777 &0.1685  &0.0032$\pm$0.0005  &14.70$\pm$0.49 &0.20$\pm$0.00 &321$\pm$51 &0.013$\pm$0.036 &14.64$\pm$0.50 &9.95$\pm$0.69\\
 \enddata
\end{deluxetable}

\section{Conclusion and discussion}
This paper investigates the potential importance of binary evolution
in UV-optical spectral fitting of early-type galaxies, via a new
spectral fitting code (BS2fit). Both the results of mock and real
galaxies show that stellar population models including binary
evolution will give different estimates for galaxy parameters such
as young-component age and dust extinction. The difference between
bsSP and ssSP-fitted young-component ages ($t_{\rm 2}$) can be as
large as 7\,Gyr (mock galaxies) or 10\,Gyr (real galaxies), while
the optical depth ($\tau_{\rm V}$) can be different by about 0.15 if
ssSP-fitted $\tau_{\rm V}$ is low ($<$ 0.25). Our results suggest
that binary evolution can possibly play an important role in
spectral fitting of early-type galaxies. It is therefore necessary
to take binaries into account in spectral synthesis studies. Because
stars of most galaxies are possibly formed in more than one star
bursts, and SFHs supply unique information for studying galaxy
formation and evolution, it is crucial to include the effects of
binaries in stellar population models when studying the SFHs of
galaxies. In addition, this study shows that bsSP models measure
larger optical depth for galactic dust, because bsSPs are more
luminous than ssSPs in UV band. This is possibly helpful for
explaining why some minus extinction values are needed in some
spectral fittings based on ssSP models (e.g.,
\citealt{2005MNRAS.358..363C}). Furthermore, although parameters
such as old-component age and stellar velocity dispersion, which are
derived from bsSP and ssSP models, are roughly consistent, the
results for some specific galaxies are different a lot (e.g., 8\,Gyr
for $t_{\rm 1}$ and 100 km~s$^{-1}$ for $\sigma_*$). This implies
that binary evolution may affect more studies about galaxies.

A fixed binary fraction (50\%) is taken for all bsSPs, although the
real fraction may be different from a galaxy to another. However,
our results are potentially useful, because 50\% is a typical binary
fraction of our Galaxy, and this fraction can fit the spectra of
most galaxies. However, the assumption of separation of binary
components may have obvious effects on the final results, although
other adjustable parameters affect the results slightly
\citep{2002MNRAS.329..897H,2012MNRAS.424..874L}. On another side,
although it is possible that UV-upturn is caused mainly by binary
evolution, some single extreme horizontal branch (EHB) stars can
also affect the UV flux of early-type galaxies (e.g.,
\citealt{2000ApJ...532..308B}). Thus our results actually show the
maximum effects of binaries on spectral fittings of early-type
galaxies. In addition, a simple assumption for SFH was used to model
the CSPs of early-type galaxies. This is possibly reasonable for
most galaxies, but it is interesting to do some researches via
taking different SFHs. Furthermore, the stellar masses of
populations were measured by the flux at 5500\,${\rm \AA}$ in this
work. If stellar mass is determined at other wavelengths, e.g.,
4000\,${\rm \AA}$, the bsSP and ssSP-fitted stellar masses may be
different more. Moreover, we used a rapid spectral fitting technique
in BS2fit code because of the limitation of our computation ability.
This leads to some uncertainties in the results. Although this does
not affect the results of this work too much, it is necessary to
make the code more accurate and check the results again.

\acknowledgments

We greatly acknowledge Profs. Pavel Kroupa, Xiangdong Li, Qiusheng
Gu, and Shiyin Shen for suggestions and discussions. This work is
supported by Chinese National Science Foundation (Grant No.
10963001, 11203005), Yunnan Science Foundation (2009CD093), and
Chinese Postdoctral Science Foundation. We also thank the support of
Sino-German Center (GZ585) and K.C. Wong Education Foundation.
\clearpage


\begin{thebibliography}{74}
\expandafter\ifx\csname
natexlab\endcsname\relax\def\natexlab#1{#1}\fi

\bibitem[{{Allard} {et~al.}(1994){Allard}, {Wesemael}, {Fontaine}, {Bergeron},
  \& {Lamontagne}}]{1994AJ....107.1565A}
{Allard}, F., {Wesemael}, F., {Fontaine}, G., {Bergeron}, P., \&
{Lamontagne},
  R. 1994, \aj, 107, 1565

\bibitem[{{Aznar Cuadrado} \& {Jeffery}(2001)}]{2001A&A...368..994A}
{Aznar Cuadrado}, R., \& {Jeffery}, C.~S. 2001, \aap, 368, 994

\bibitem[{{Bressan} {et~al.}(1994){Bressan}, {Chiosi}, \&
  {Fagotto}}]{1994ApJS...94...63B}
{Bressan}, A., {Chiosi}, C., \& {Fagotto}, F. 1994, \apjs, 94, 63

\bibitem[{{Brinchmann}(2010)}]{2010IAUS..262....3B}
{Brinchmann}, J. 2010, in IAU Symposium, Vol. 262, IAU Symposium,
ed.
  G.~{Bruzual} \& S.~{Charlot}, 3--12

\bibitem[{{Brown} {et~al.}(2000){Brown}, {Bowers}, {Kimble}, {Sweigart}, \&
  {Ferguson}}]{2000ApJ...532..308B}
{Brown}, T.~M., {Bowers}, C.~W., {Kimble}, R.~A., {Sweigart}, A.~V.,
\&
  {Ferguson}, H.~C. 2000, \apj, 532, 308

\bibitem[{{Burstein} \& {Heiles}(1982)}]{1982AJ.....87.1165B}
{Burstein}, D., \& {Heiles}, C. 1982, \aj, 87, 1165

\bibitem[{{Cardelli} {et~al.}(1989){Cardelli}, {Clayton}, \&
  {Mathis}}]{1989ApJ...345..245C}
{Cardelli}, J.~A., {Clayton}, G.~C., \& {Mathis}, J.~S. 1989, \apj,
345, 245

\bibitem[{{Carney} {et~al.}(2005){Carney}, {Aguilar}, {Latham}, \&
  {Laird}}]{2005AJ....129.1886C}
{Carney}, B.~W., {Aguilar}, L.~A., {Latham}, D.~W., \& {Laird},
J.~B. 2005,
  \aj, 129, 1886

\bibitem[{{Chabrier}(2003)}]{Chabrier:2003}
{Chabrier}, G. 2003, {ApJ}, 586, L133

\bibitem[{{Charlot} \& {Fall}(2000)}]{2000ApJ...539..718C}
{Charlot}, S., \& {Fall}, S.~M. 2000, \apj, 539, 718

\bibitem[{{Chen} {et~al.}(2012){Chen}, {Kauffmann}, {Tremonti}, {White},
  {Heckman}, {Kova{\v c}}, {Bundy}, {Chisholm}, {Maraston}, {Schneider},
  {Bolton}, {Weaver}, \& {Brinkmann}}]{2012MNRAS.421..314C}
{Chen}, Y.-M., {Kauffmann}, G., {Tremonti}, C.~A., {White}, S.,
{Heckman},
  T.~M., {Kova{\v c}}, K., {Bundy}, K., {Chisholm}, J., {Maraston}, C.,
  {Schneider}, D.~P., {Bolton}, A.~S., {Weaver}, B.~A., \& {Brinkmann}, J.
  2012, \mnras, 421, 314

\bibitem[{{Chilingarian} {et~al.}(2007){Chilingarian}, {Prugniel},
  {Sil'Chenko}, \& {Koleva}}]{2007IAUS..241..175C}
{Chilingarian}, I., {Prugniel}, P., {Sil'Chenko}, O., \& {Koleva},
M. 2007, in
  IAU Symposium, Vol. 241, IAU Symposium, ed. {A.~Vazdekis \& R.~F.~Peletier},
  175--176

\bibitem[{{Cid Fernandes} {et~al.}(2004){Cid Fernandes}, {Gu}, {Melnick},
  {Terlevich}, {Terlevich}, {Kunth}, {Rodrigues Lacerda}, \&
  {Joguet}}]{2004MNRAS.355..273C}
{Cid Fernandes}, R., {Gu}, Q., {Melnick}, J., {Terlevich}, E.,
{Terlevich}, R.,
  {Kunth}, D., {Rodrigues Lacerda}, R., \& {Joguet}, B. 2004, \mnras, 355, 273

\bibitem[{{Cid Fernandes} {et~al.}(2005){Cid Fernandes}, {Mateus}, {Sodr{\'e}},
  {Stasi{\'n}ska}, \& {Gomes}}]{2005MNRAS.358..363C}
{Cid Fernandes}, R., {Mateus}, A., {Sodr{\'e}}, L., {Stasi{\'n}ska},
G., \&
  {Gomes}, J.~M. 2005, \mnras, 358, 363

\bibitem[{{Colless} {et~al.}(2001){Colless}, {Dalton}, {Maddox}, {Sutherland},
  {Norberg}, {Cole}, {Bland-Hawthorn}, {Bridges}, {Cannon}, {Collins}, {Couch},
  {Cross}, {Deeley}, {De Propris}, {Driver}, {Efstathiou}, {Ellis}, {Frenk},
  {Glazebrook}, {Jackson}, {Lahav}, {Lewis}, {Lumsden}, {Madgwick}, {Peacock},
  {Peterson}, {Price}, {Seaborne}, \& {Taylor}}]{2001MNRAS.328.1039C}
{Colless}, M., {Dalton}, G., {Maddox}, S., {Sutherland}, W.,
{Norberg}, P.,
  {Cole}, S., {Bland-Hawthorn}, J., {Bridges}, T., {Cannon}, R., {Collins}, C.,
  {Couch}, W., {Cross}, N., {Deeley}, K., {De Propris}, R., {Driver}, S.~P.,
  {Efstathiou}, G., {Ellis}, R.~S., {Frenk}, C.~S., {Glazebrook}, K.,
  {Jackson}, C., {Lahav}, O., {Lewis}, I., {Lumsden}, S., {Madgwick}, D.,
  {Peacock}, J.~A., {Peterson}, B.~A., {Price}, I., {Seaborne}, M., \&
  {Taylor}, K. 2001, \mnras, 328, 1039

\bibitem[{{da Costa} {et~al.}(1991){da Costa}, {Pellegrini}, {Davis},
  {Meiksin}, {Sargent}, \& {Tonry}}]{1991ApJS...75..935D}
{da Costa}, L.~N., {Pellegrini}, P.~S., {Davis}, M., {Meiksin}, A.,
{Sargent},
  W.~L.~W., \& {Tonry}, J.~L. 1991, \apjs, 75, 935

\bibitem[{{Denicol{\'o}} {et~al.}(2005){Denicol{\'o}}, {Terlevich},
  {Terlevich}, {Forbes}, {Terlevich}, \& {Carrasco}}]{2005MNRAS.356.1440D}
{Denicol{\'o}}, G., {Terlevich}, R., {Terlevich}, E., {Forbes},
D.~A.,
  {Terlevich}, A., \& {Carrasco}, L. 2005, \mnras, 356, 1440

\bibitem[{{Duquennoy} \& {Mayor}(1991)}]{1991AA...248..485D}
{Duquennoy}, A., \& {Mayor}, M. 1991, \aap, 248, 485

\bibitem[{{Eggleton}(2006)}]{Eggleton:2006}
{Eggleton}, P. 2006, Evolutionary Process in Binary and Multiple
Stars,
  Cambridge University Press

\bibitem[{{Ferguson} {et~al.}(1984){Ferguson}, {Green}, \&
  {Liebert}}]{1984ApJ...287..320F}
{Ferguson}, D.~H., {Green}, R.~F., \& {Liebert}, J. 1984, \apj, 287,
320

\bibitem[{{Fischer} \& {Marcy}(1992)}]{1992ApJ...396..178F}
{Fischer}, D.~A., \& {Marcy}, G.~W. 1992, \apj, 396, 178

\bibitem[{{Goldberg} \& {Mazeh}(1994)}]{Goldberg:1994}
{Goldberg}, D., \& {Mazeh}, T. 1994, {A\&A}, 282, 801

\bibitem[{{Han} {et~al.}(1995){Han}, {Podsiadlowski}, \& {Eggleton}}]{Han:1995}
{Han}, Z., {Podsiadlowski}, P., \& {Eggleton}, P.~P. 1995, {MNRAS},
272, 800

\bibitem[{{Han} {et~al.}(2007){Han}, {Podsiadlowski}, \&
  {Lynas-Gray}}]{2007MNRAS.380.1098H}
{Han}, Z., {Podsiadlowski}, P., \& {Lynas-Gray}, A.~E. 2007, \mnras,
380, 1098

\bibitem[{{Heavens} {et~al.}(2000){Heavens}, {Jimenez}, \&
  {Lahav}}]{2000MNRAS.317..965H}
{Heavens}, A.~F., {Jimenez}, R., \& {Lahav}, O. 2000, \mnras, 317,
965

\bibitem[{{Hern{\'a}ndez} \& {Bruzual}(2011)}]{2011RMxAC..40..277H}
{Hern{\'a}ndez}, F.~C., \& {Bruzual}, G. 2011, in Revista Mexicana
de
  Astronomia y Astrofisica Conference Series, Vol.~40, Revista Mexicana de
  Astronomia y Astrofisica Conference Series, 277--277

\bibitem[{{Hjellming} \& {Webbink}(1987)}]{1987ApJ...318..794H}
{Hjellming}, M.~S., \& {Webbink}, R.~F. 1987, \apj, 318, 794

\bibitem[{{Huchra} {et~al.}(1999){Huchra}, {Vogeley}, \&
  {Geller}}]{1999ApJS..121..287H}
{Huchra}, J.~P., {Vogeley}, M.~S., \& {Geller}, M.~J. 1999, \apjs,
121, 287

\bibitem[{{Hurley} {et~al.}(2002){Hurley}, {Tout}, \&
  {Pols}}]{2002MNRAS.329..897H}
{Hurley}, J.~R., {Tout}, C.~A., \& {Pols}, O.~R. 2002, \mnras, 329,
897

\bibitem[{{Kinney} {et~al.}(1996){Kinney}, {Calzetti}, {Bohlin}, {McQuade},
  {Storchi-Bergmann}, \& {Schmitt}}]{1996ApJ...467...38K}
{Kinney}, A.~L., {Calzetti}, D., {Bohlin}, R.~C., {McQuade}, K.,
  {Storchi-Bergmann}, T., \& {Schmitt}, H.~R. 1996, \apj, 467, 38

\bibitem[{{Koleva} {et~al.}(2009){Koleva}, {Prugniel}, {Bouchard}, \&
  {Wu}}]{2009A&A...501.1269K}
{Koleva}, M., {Prugniel}, P., {Bouchard}, A., \& {Wu}, Y. 2009,
\aap, 501, 1269

\bibitem[{{Koleva} {et~al.}(2008){Koleva}, {Prugniel}, {Ocvirk}, {Le Borgne},
  \& {Soubiran}}]{2008MNRAS.385.1998K}
{Koleva}, M., {Prugniel}, P., {Ocvirk}, P., {Le Borgne}, D., \&
{Soubiran}, C.
  2008, \mnras, 385, 1998

\bibitem[{{Koribalski} {et~al.}(2004){Koribalski}, {Staveley-Smith}, {Kilborn},
  {Ryder}, {Kraan-Korteweg}, {Ryan-Weber}, {Ekers}, {Jerjen}, {Henning},
  {Putman}, {Zwaan}, {de Blok}, {Calabretta}, {Disney}, {Minchin}, {Bhathal},
  {Boyce}, {Drinkwater}, {Freeman}, {Gibson}, {Green}, {Haynes}, {Juraszek},
  {Kesteven}, {Knezek}, {Mader}, {Marquarding}, {Meyer}, {Mould}, {Oosterloo},
  {O'Brien}, {Price}, {Sadler}, {Schr{\"o}der}, {Stewart}, {Stootman}, {Waugh},
  {Warren}, {Webster}, \& {Wright}}]{2004AJ....128...16K}
{Koribalski}, B.~S., {Staveley-Smith}, L., {Kilborn}, V.~A.,
{Ryder}, S.~D.,
  {Kraan-Korteweg}, R.~C., {Ryan-Weber}, E.~V., {Ekers}, R.~D., {Jerjen}, H.,
  {Henning}, P.~A., {Putman}, M.~E., {Zwaan}, M.~A., {de Blok}, W.~J.~G.,
  {Calabretta}, M.~R., {Disney}, M.~J., {Minchin}, R.~F., {Bhathal}, R.,
  {Boyce}, P.~J., {Drinkwater}, M.~J., {Freeman}, K.~C., {Gibson}, B.~K.,
  {Green}, A.~J., {Haynes}, R.~F., {Juraszek}, S., {Kesteven}, M.~J., {Knezek},
  P.~M., {Mader}, S., {Marquarding}, M., {Meyer}, M., {Mould}, J.~R.,
  {Oosterloo}, T., {O'Brien}, J., {Price}, R.~M., {Sadler}, E.~M.,
  {Schr{\"o}der}, A., {Stewart}, I.~M., {Stootman}, F., {Waugh}, M., {Warren},
  B.~E., {Webster}, R.~L., \& {Wright}, A.~E. 2004, \aj, 128, 16

\bibitem[{{Lada} \& {Lada}(2003)}]{2003ARAA..41...57L}
{Lada}, C.~J., \& {Lada}, E.~A. 2003, \araa, 41, 57

\bibitem[{{Lee}(1994)}]{1994ApJ...430L.113L}
{Lee}, Y.-W. 1994, \apjl, 430, L113

\bibitem[{{Lejeune} {et~al.}(1997){Lejeune}, {Cuisinier}, \&
  {Buser}}]{Lejeune:1997}
{Lejeune}, T., {Cuisinier}, F., \& {Buser}, R. 1997, {A\&AS}, 125,
229

\bibitem[{{Lejeune} {et~al.}(1998){Lejeune}, {Cuisinier}, \&
  {Buser}}]{Lejeune:1998}
---. 1998, {A\&AS}, 130, 65

\bibitem[{{Li} \& {Han}(2007)}]{2007A&A...471..795L}
{Li}, Z., \& {Han}, Z. 2007, \aap, 471, 795

\bibitem[{{Li} \& {Han}(2008{\natexlab{a}})}]{2008MNRAS.387..105L}
---. 2008{\natexlab{a}}, \mnras, 387, 105

\bibitem[{{Li} \& {Han}(2008{\natexlab{b}})}]{2008ApJ...685..225L}
---. 2008{\natexlab{b}}, \apj, 685, 225

\bibitem[{{Li} \& {Han}(2008{\natexlab{c}})}]{2008IAUS..252..359L}
{Li}, Z., \& {Han}, Z. 2008{\natexlab{c}}, in IAU Symposium, Vol.
252, IAU
  Symposium, ed. L.~{Deng} \& K.~L. {Chan}, 359--364

\bibitem[{{Li} {et~al.}(2012{\natexlab{a}}){Li}, {Zhang}, \&
  {Liu}}]{2012MNRAS.424..874L}
{Li}, Z., {Zhang}, L., \& {Liu}, J. 2012{\natexlab{a}}, \mnras, 424,
874

\bibitem[{{Li} {et~al.}(2010){Li}, {Mao}, {Li}, {Li}, \&
  {Li}}]{2010RAA....10..135L}
{Li}, Z.-M., {Mao}, C.-Y., {Li}, R.-H., {Li}, R.-X., \& {Li}, M.-C.
2010,
  Research in Astronomy and Astrophysics, 10, 135

\bibitem[{{Li} {et~al.}(2012{\natexlab{b}}){Li}, {Mao}, {Zhang}, \&
  {Chen}}]{2012ApJL}
{Li}, Z.-M., {Mao}, C.-Y., {Zhang}, Q., \& {Chen}, L.
2012{\natexlab{b}},
  Astrophysical Journal Letters, 761, 22

\bibitem[{{Li} {et~al.}(2006){Li}, {Zhang}, \& {Han}}]{2006ChJAA...6..669L}
{Li}, Z.-M., {Zhang}, F.-H., \& {Han}, Z.-W. 2006, \cjaa, 6, 669

\bibitem[{{Lu} {et~al.}(2006){Lu}, {Zhou}, {Wang}, {Wang}, {Dong}, {Zhuang}, \&
  {Li}}]{2006AJ....131..790L}
{Lu}, H., {Zhou}, H., {Wang}, J., {Wang}, T., {Dong}, X., {Zhuang},
Z., \&
  {Li}, C. 2006, \aj, 131, 790

\bibitem[{{Marks} \& {Kroupa}(2011)}]{2011MNRAS.417.1702M}
{Marks}, M., \& {Kroupa}, P. 2011, \mnras, 417, 1702

\bibitem[{{Marks} \& {Kroupa}(2012)}]{2012AA...543A...8M}
---. 2012, \aap, 543, A8

\bibitem[{{Maxted} {et~al.}(2001){Maxted}, {Heber}, {Marsh}, \&
  {North}}]{2001MNRAS.326.1391M}
{Maxted}, P.~f.~L., {Heber}, U., {Marsh}, T.~R., \& {North}, R.~C.
2001,
  \mnras, 326, 1391

\bibitem[{{Mayor} {et~al.}(1992){Mayor}, {Duquennoy}, {Halbwachs}, \&
  {Mermilliod}}]{1992ASPC...32...73M}
{Mayor}, M., {Duquennoy}, A., {Halbwachs}, J.-L., \& {Mermilliod},
J.-C. 1992,
  in Astronomical Society of the Pacific Conference Series, Vol.~32, IAU
  Colloq. 135: Complementary Approaches to Double and Multiple Star Research,
  ed. H.~A. {McAlister} \& W.~I. {Hartkopf}, 73

\bibitem[{{Mazeh} {et~al.}(1992){Mazeh}, {Goldberg}, {Duquennoy}, \&
  {Mayor}}]{Mazeh:1992}
{Mazeh}, T., {Goldberg}, D., {Duquennoy}, A., \& {Mayor}, M. 1992,
{ApJ}, 401,
  265

\bibitem[{{McQuade} {et~al.}(1995){McQuade}, {Calzetti}, \&
  {Kinney}}]{1995ApJS...97..331M}
{McQuade}, K., {Calzetti}, D., \& {Kinney}, A.~L. 1995, \apjs, 97,
331

\bibitem[{{O'Connell}(1999)}]{1999ARAA..37..603O}
{O'Connell}, R.~W. 1999, \araa, 37, 603

\bibitem[{{Ocvirk} {et~al.}(2006){Ocvirk}, {Pichon}, {Lan{\c c}on}, \&
  {Thi{\'e}baut}}]{2006MNRAS.365...74O}
{Ocvirk}, P., {Pichon}, C., {Lan{\c c}on}, A., \& {Thi{\'e}baut}, E.
2006,
  \mnras, 365, 74

\bibitem[{{Panter} {et~al.}(2007){Panter}, {Jimenez}, {Heavens}, \&
  {Charlot}}]{2007MNRAS.378.1550P}
{Panter}, B., {Jimenez}, R., {Heavens}, A.~F., \& {Charlot}, S.
2007, \mnras,
  378, 1550

\bibitem[{{Park} \& {Lee}(1997)}]{1997ApJ...476...28P}
{Park}, J.-H., \& {Lee}, Y.-W. 1997, \apj, 476, 28

\bibitem[{{Raghavan} {et~al.}(2010){Raghavan}, {McAlister}, {Henry}, {Latham},
  {Marcy}, {Mason}, {Gies}, {White}, \& {ten Brummelaar}}]{2010ApJS..190....1R}
{Raghavan}, D., {McAlister}, H.~A., {Henry}, T.~J., {Latham}, D.~W.,
{Marcy},
  G.~W., {Mason}, B.~D., {Gies}, D.~R., {White}, R.~J., \& {ten Brummelaar},
  T.~A. 2010, \apjs, 190, 1

\bibitem[{{Reed} \& {Stiening}(2004)}]{2004PASP..116..506R}
{Reed}, M.~D., \& {Stiening}, R. 2004, \pasp, 116, 506

\bibitem[{{Schlegel} {et~al.}(1998){Schlegel}, {Finkbeiner}, \&
  {Davis}}]{1998ApJ...500..525S}
{Schlegel}, D.~J., {Finkbeiner}, D.~P., \& {Davis}, M. 1998, \apj,
500, 525

\bibitem[{{Thejll} {et~al.}(1995){Thejll}, {Ulla}, \&
  {MacDonald}}]{1995A&A...303..773T}
{Thejll}, P., {Ulla}, A., \& {MacDonald}, J. 1995, \aap, 303, 773

\bibitem[{{Thomas} {et~al.}(2005){Thomas}, {Maraston}, {Bender}, \& {Mendes de
  Oliveira}}]{Thomas:2005}
{Thomas}, D., {Maraston}, C., {Bender}, R., \& {Mendes de Oliveira},
C. 2005,
  {ApJ}, 621, 673

\bibitem[{{Tojeiro} {et~al.}(2007){Tojeiro}, {Heavens}, {Jimenez}, \&
  {Panter}}]{2007MNRAS.381.1252T}
{Tojeiro}, R., {Heavens}, A.~F., {Jimenez}, R., \& {Panter}, B.
2007, \mnras,
  381, 1252

\bibitem[{{Tojeiro} {et~al.}(2009){Tojeiro}, {Wilkins}, {Heavens}, {Panter}, \&
  {Jimenez}}]{2009ApJS..185....1T}
{Tojeiro}, R., {Wilkins}, S., {Heavens}, A.~F., {Panter}, B., \&
{Jimenez}, R.
  2009, \apjs, 185, 1

\bibitem[{{Tout} {et~al.}(1997){Tout}, {Aarseth}, {Pols}, \&
  {Eggleton}}]{1997MNRAS.291..732T}
{Tout}, C.~A., {Aarseth}, S.~J., {Pols}, O.~R., \& {Eggleton}, P.~P.
1997,
  \mnras, 291, 732

\bibitem[{{Tremonti} {et~al.}(2004){Tremonti}, {Heckman}, {Kauffmann},
  {Brinchmann}, {Charlot}, {White}, {Seibert}, {Peng}, {Schlegel}, {Uomoto},
  {Fukugita}, \& {Brinkmann}}]{2004ApJ...613..898T}
{Tremonti}, C.~A., {Heckman}, T.~M., {Kauffmann}, G., {Brinchmann},
J.,
  {Charlot}, S., {White}, S.~D.~M., {Seibert}, M., {Peng}, E.~W., {Schlegel},
  D.~J., {Uomoto}, A., {Fukugita}, M., \& {Brinkmann}, J. 2004, \apj, 613, 898

\bibitem[{{Tully} {et~al.}(2008){Tully}, {Shaya}, {Karachentsev}, {Courtois},
  {Kocevski}, {Rizzi}, \& {Peel}}]{2008ApJ...676..184T}
{Tully}, R.~B., {Shaya}, E.~J., {Karachentsev}, I.~D., {Courtois},
H.~M.,
  {Kocevski}, D.~D., {Rizzi}, L., \& {Peel}, A. 2008, \apj, 676, 184

\bibitem[{{Ulla} \& {Thejll}(1998)}]{1998A&AS..132....1U}
{Ulla}, A., \& {Thejll}, P. 1998, \aaps, 132, 1

\bibitem[{{Walcher} {et~al.}(2006){Walcher}, {B{\"o}ker}, {Charlot}, {Ho},
  {Rix}, {Rossa}, {Shields}, \& {van der Marel}}]{2006ApJ...649..692W}
{Walcher}, C.~J., {B{\"o}ker}, T., {Charlot}, S., {Ho}, L.~C.,
{Rix}, H.-W.,
  {Rossa}, J., {Shields}, J.~C., \& {van der Marel}, R.~P. 2006, \apj, 649, 692

\bibitem[{{Walcher} {et~al.}(2011){Walcher}, {Groves}, {Budav{\'a}ri}, \&
  {Dale}}]{2011Ap&SS.331....1W}
{Walcher}, J., {Groves}, B., {Budav{\'a}ri}, T., \& {Dale}, D. 2011,
\apss,
  331, 1

\bibitem[{{Wegner} {et~al.}(2003){Wegner}, {Bernardi}, {Willmer}, {da Costa},
  {Alonso}, {Pellegrini}, {Maia}, {Chaves}, \&
  {Rit{\'e}}}]{2003AJ....126.2268W}
{Wegner}, G., {Bernardi}, M., {Willmer}, C.~N.~A., {da Costa},
L.~N., {Alonso},
  M.~V., {Pellegrini}, P.~S., {Maia}, M.~A.~G., {Chaves}, O.~L., \& {Rit{\'e}},
  C. 2003, \aj, 126, 2268

\bibitem[{{Westera} {et~al.}(2002){Westera}, {Lejeune}, {Buser}, {Cuisinier},
  \& {Bruzual}}]{Westera:2002}
{Westera}, P., {Lejeune}, T., {Buser}, R., {Cuisinier}, F., \&
{Bruzual}, G.
  2002, {A\&A}, 381, 524

\bibitem[{{Williams} {et~al.}(2001){Williams}, {McGraw}, {Mason}, \&
  {Grashuis}}]{2001PASP..113..944W}
{Williams}, T., {McGraw}, J.~T., {Mason}, P.~A., \& {Grashuis}, R.
2001, \pasp,
  113, 944

\bibitem[{{Yi} {et~al.}(1997){Yi}, {Demarque}, \& {Kim}}]{1997ApJ...482..677Y}
{Yi}, S., {Demarque}, P., \& {Kim}, Y.-C. 1997, \apj, 482, 677

\bibitem[{{Yi} {et~al.}(2005){Yi}, {Yoon}, {Kaviraj}, {Deharveng}, {Rich},
  {Salim}, {Boselli}, {Lee}, {Ree}, {Sohn}, {Rey}, {Lee}, {Rhee}, {Bianchi},
  {Byun}, {Donas}, {Friedman}, {Heckman}, {Jelinsky}, {Madore}, {Malina},
  {Martin}, {Milliard}, {Morrissey}, {Neff}, {Schiminovich}, {Siegmund},
  {Small}, {Szalay}, {Jee}, {Kim}, {Barlow}, {Forster}, {Welsh}, \&
  {Wyder}}]{Yi:2005}
{Yi}, S.~K., {Yoon}, S.-J., {Kaviraj}, S., {Deharveng}, J.-M.,
{Rich}, R.~M.,
  {Salim}, S., {Boselli}, A., {Lee}, Y.-W., {Ree}, C.~H., {Sohn}, Y.-J., {Rey},
  S.-C., {Lee}, J.-W., {Rhee}, J., {Bianchi}, L., {Byun}, Y.-I., {Donas}, J.,
  {Friedman}, P.~G., {Heckman}, T.~M., {Jelinsky}, P., {Madore}, B.~F.,
  {Malina}, R., {Martin}, D.~C., {Milliard}, B., {Morrissey}, P., {Neff}, S.,
  {Schiminovich}, D., {Siegmund}, O., {Small}, T., {Szalay}, A.~S., {Jee},
  M.~J., {Kim}, S.-W., {Barlow}, T., {Forster}, K., {Welsh}, B., \& {Wyder},
  T.~K. 2005, {ApJ}, 619, L111
  
\end{thebibliography}

\end{document}